\documentclass[aps,prl,twocolumn,floatfix]{revtex4-1}
\usepackage{amsmath}
\usepackage{amsfonts}
\usepackage{graphicx}
\usepackage{color}
\usepackage{bm}
\newcommand{\TTT}{\mathcal{T}}
\newcommand{\PPP}{\mathcal{P}}
\newcommand{\CCC}{\mathcal{C}}

\begin{document}

\title{Local Adiabatic Mixing of Kramers Pairs of Majorana Bound States}
\author{Konrad W\"olms,$^1$ Ady Stern,$^2$ and Karsten Flensberg$^1$}
\affiliation{$^1$Center for Quantum Devices, Niels Bohr Institute,
University of Copenhagen, Universitetsparken 5, 2100 Copenhagen, Denmark\\
$^2$Department of Condensed Matter Physics, Weizmann Institute of Science, Rehovot 76100, Israel}

\date{\today}

\begin{abstract}
    We consider Kramers pairs of Majorana bound states under adiabatic time evolution.  This is
    important for the prospects of using such bound states as parity qubits. We show that local
    adiabatic perturbations can cause a rotation in  the space spanned by the Kramers pair. Hence the quantum information is unprotected against local perturbations, in contrast to the case of single localized Majorana bound states in systems with broken time reversal symmetry. We give an analytical and a numerical example for such a rotation, and specify sufficient conditions under which a rotation is avoided. We give a general scheme for determining when these conditions are satisfied, and exemplify it with a general model of a quasi 1D time reversal symmetric topological superconductor.
\end{abstract}

\pacs{71.10.Pm, 74.45.+c, 74.78.Na}
\maketitle

Majorana bound states (or Majorana Fermions) in condensed matter systems have been the subject of a
large research effort in the last few years. Among other reasons, this effort has been motivated by
a number of recent proposals for feasible experimental systems hosting Majorana bound states
(MBS)\cite{AliceaReview,BeenakkerReview,LeijnseReview}, and by their relevance to topological
quantum computation\cite{Nayak2008}. In superconducting systems, a MBS describes a localized zero
energy solution of the Bogoliubov-deGennes (BdG) equation. Such a solution constitutes "half" a
Fermion, and two such solutions span a fermionic mode, of two states. The zero energy solutions of
the BdG equations signify a degeneracy of the superconducting many-body ground state, defining a degenerate subspace within
which manipulations are possible through adiabatic variation of the Hamiltonian. A particularly
interesting set of manipulations is braiding of the positions of the MBSs (while maintaining the
degeneracy of the ground state), which constitutes a set of non-abelian operations
referred to as gates. If the MBS are all spatially separated from one another, these gates are expected to be topologically
protected. If the distance between the MBS, $L$, is much larger than their localization
length $\xi$, the unitary transformation associated with their braiding is topologically stable,
which means that corrections are exponentially small in the ratio $L/\xi$. As such, they are exponentially small in
$E_g$, the energy gap of the superconductor.

The isolation of single localized zero
energy solutions requires a system where time reversal symmetry (TRS) is broken, since under
TRS the solutions of the BdG equation form degenerate Kramers pairs, and isolation of an odd
number of localized solutions is impossible.
Recently, there has been a large interest in topological superconductors respecting  TRS, i.e. in
absence of magnetic fields (or spontaneously broken TRS). A number of proposals for systems in
hybrid materials and
structures\cite{Wong2012,Zhang2013,Nakosai2013,Nakosai2012,Deng2012,Keselman2013,Gaidamauskas2014,Haim2013}
have been put forward. Such systems accommodate Kramers pairs of MBSs, and
therefore do not allow for braiding of single Majorana fermions. Since these systems allow for
braiding of Kramers pairs of MBS, the question of the possible protection of quantum information
in Kramers pairs of MBSs arises. It has been suggested to use braiding of such MBS pairs for
topological quantum computation in a similar way to isolated
MBS \cite{Liu2014}.

In this letter we show that MBS Kramers pairs are not protected against local adiabatic
changes of the Hamiltonian even if the parameter dependent  Hamiltonian is TRS for every
value of the parameters. We also give a characterization of adiabatic changes that maintain protection,
which could form the basis of engineering systems with some degree of protection. Without this
protection, Majorana qubits in TRS topological superconductors are prone to the decoherence by local
perturbation similar to non-topological qubits.

We consider a finite DIII wire, i.e. a 1D system with time reversal symmetry $\TTT^2=-1$ and
particle-hole symmetry, in the topological phase. There is a Kramers pair of MBSs, $\gamma_L$ and
$\tilde\gamma_L=\TTT\gamma_L\TTT^{-1}$, located on the left end of the wire, and another pair,
$\gamma_R$ and $\tilde\gamma_R$, located on the right end of the wire.  Let us recall the defining
mathematical properties of Majorana operators, which are $\gamma_i^\dagger=\gamma_i$ and
$\left\{\gamma_i,\gamma_j  \right\}=2\delta_{ij}$, where $i$ and $j$ are generic indexes, which may
run over left and right and time reversed partners.  We define local mixing of a Kramers pair as
unitary transformations that only involve operators of one Kramers pair, e.g.
$\gamma_R,\tilde\gamma_R$. Such transformations take the form
$U=e^{i\delta}e^{\varphi\gamma_R\tilde\gamma_R/2}$. We refer to this as mixing because the transformation rotates the corresponding Majoranas: $U\gamma_R
U^\dagger=\gamma_R\cos\varphi+\tilde\gamma_R\sin\varphi$ and $U\tilde\gamma_R
U^\dagger=\tilde\gamma_R\cos\varphi-\gamma_R\sin\varphi$.  Before we show that adiabatic processes
may cause local mixing with a value of $\varphi$ that is generally non-quantized we discuss an implication for the ground state of the system.  We set $\delta=0$
because it is just an overall phase. The ground state subspace can be described by forming the
non-local fermions $c=\tfrac 1 2(\gamma_R + i\gamma_L)$ and $\tilde c=\tfrac 1 2(\tilde\gamma_R
-i\tilde\gamma_L)$, such that $\TTT c\TTT^{-1}=\tilde c$. We label the ground states according to
the eigenvalues of the occupation number operators that are associated with $c$ and $\tilde c$.  If the transformation $U$
is applied to the state  $|00\rangle$, whose first and second entries correspond to the occupations of the $c, {\tilde c}$ fermions, respectively, its operation yields $U|00\rangle=\cos\tfrac \varphi 2|00\rangle +
\sin\tfrac \varphi 2|11\rangle$. This shows that a finite $\varphi$ mixes the fermion parity among
the initial basis choice of time reversed partners. Interestingly, the states $|00\rangle$ and $|11\rangle$ are only degenerate because of the topology of the system and not due to Kramers theorem. Therefore, even though the mixing is generated by a Kramers pair of Majorana fermions, it acts on degenerate states that are unique to systems in a topological phase.

In an adiabatic process a set of parameters, $\bm{\eta}$, varies slowly in time. To each parameter value corresponds a Hamiltonian $H(\bm{\eta})$, which we assume to be time reversal symmetric
for all $\bm{\eta}$: $[H(\bm{\eta}),\TTT]=0$. $H(\bm{\eta})$ is therefore always in the DIII symmetry class and has Kramers pairs of MBSs. We further assume, for simplicity,  that the variation of $\eta$ changes the Hamiltonian only at the right side of the wire. Then, $\gamma_L$ and $\tilde\gamma_L$ do not vary with $\bm\eta$. Thus, everything that follows will only use $\gamma_R$ and $\tilde{\gamma}_R$(and their linear combinations), for which we drop the subscript $R$. For each $\bm{\eta}$ we choose a particular Kramers pair of MBS $\gamma_{\bm{\eta}}$ and $\tilde\gamma_{\bm\eta}$. This choice is not unique, because there are linear
combinations of $\gamma_{\bm\eta}$ and $\tilde\gamma_{\bm{\eta}}$ that also form a Kramers pair of MBSs.
Generally, any adiabatically time-evolved Majorana operator on the right end of the wire, $\Gamma(t)$, that at $t=0$ commuted
with the Hamiltonian, can be expressed as a linear combination of $\gamma_{\bm{\eta}}$ and
$\tilde\gamma_{\bm{\eta}}$:
\begin{equation}
    \Gamma(t)=a(t)\gamma_{\bm{\eta}(t)}+b(t)\tilde\gamma_{\bm{\eta}(t)}.
    \label{eqn:MajoranaGeneral}
\end{equation}
Being initially a Majorana operator, it evolves into a Majorana operator at all later times. For $\Gamma(t)$ in Eq.~\eqref{eqn:MajoranaGeneral} to be self-adjoint, $a$ and $b$ have to be real. In addition for $\Gamma(t)$
to square to 1, $a$ and $b$ have to satisfy $a^2+b^2=1$. Therefore we parameterize $a$ and $b$ by trigonometric functions: $a(t)=\cos(\varphi(t)-\varphi_0)$ and $b(t)=\sin(\varphi(t)-\varphi_0)$. There is only
one free parameter , $\varphi$, therefore all possible processes form an Abelian group. Because the operator $\Gamma(t)$ is at zero energy during the adiabatic process it always commutes with the Hamiltonian,
which determines the time-evolution: $\tfrac {\mathrm{d}} {\mathrm{d}t}\Gamma(t)=i\left[\Gamma(t),H(t) \right]=0$.
It follows that $\big\{ \Gamma(t),\tfrac {\mathrm{d}} {\mathrm{d}t}\tilde\Gamma(t) \big\}=0$
and explicitly evaluating this anti-commutator\cite{foot1} one finds
\begin{align}
    0 &=  \big\{ \Gamma(t),\tfrac {\mathrm{d}}{\mathrm{d}t}\tilde\Gamma(t) \big\}
    \nonumber \\
    &= -2\dot\varphi +\left\{ \gamma_{\bm{\eta}(t)},\dot{\tilde\gamma}_{\bm{\eta}(t)} \right\}.
    \label{determinationVarphiDerivative}
\end{align}
This gives the important equation $\varphi(t)=\tfrac 1 2\int^t\left\{
\gamma_{\bm{\eta}(t')},\dot{\tilde\gamma}_{\bm{\eta}(t')} \right\}\mathrm{d}t'$,
which can be rewritten to be independent of time: $\varphi=\tfrac 1
2\int_\mathcal{W}\left\{\gamma_{\bm{\eta}},\nabla_{\bm{\eta}}\tilde{\gamma}_{\bm\eta}
\right\}\mathrm{d}\bm{\eta}$, where $\mathcal{W}$ is the path in parameter space traversed by
$\bm{\eta}(t)$. If the path is closed, the result will not depend on our initial choices of
$\gamma_{\bm{\eta}}$ and $\tilde\gamma_{\bm{\eta}}$ and
we can write the formula for $\varphi$ in the standard way for geometrical phases as
\begin{subequations}
    \begin{align}
        \varphi&=\oint_{\mathcal{W}}\bm{A}\mathrm{d}\bm{\eta},
        \label{eqn:BerryPhase}
        \\
        \bm{A} &=\tfrac 1 2\left\{ \gamma_{\bm{\eta}},\nabla_{\bm{\eta}} \tilde{\gamma}_{\bm{\eta}}
    \right\},
        \label{eqn:BerryPotential}
    \end{align}
    \label{eqn:BerryPhaseAndPotential}
\end{subequations}
where $\bm{A}$ is the corresponding Berry potential. By applying stokes theorem, we can obtain the
Berry curvature which is independent of our initial choices. Its components are given by
$\Omega_{\eta_i\eta_j}=\partial_{\eta_i}A_{\eta_j}-\partial_{\eta_j}A_{\eta_i}$.

\begin{figure}
    \includegraphics[]{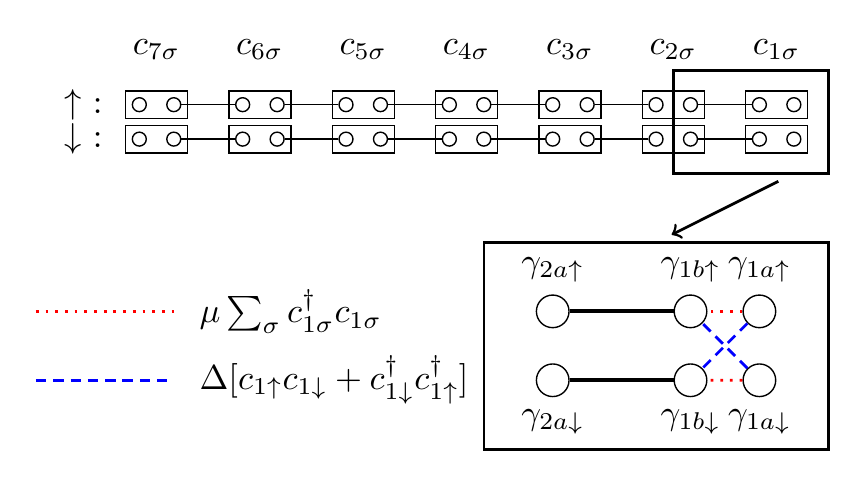}
    \caption{DIII model built out of two Kitaev chains at the special point, where the Majorana
        fermions are perfectly localized. The rectangles correspond to the original electronic
        states, and the circles correspond to their decomposition into Majorana operators according to
        $c_{i\sigma}=\tfrac 1 2 \left( \gamma_{ia\sigma} + i\gamma_{ib\sigma} \right)$. Lines indicate
        couplings.  Our model consists  of the indicated end of the wire with one Kramers
        pair of Majorana fermions and one Kramers pair of bulk states.  Furthermore it is
        illustrated how a local chemical potential and $s$-wave pairing couple the Majorana operators in our
        model. Importantly, the $s$-wave pairing couples the two initial
        Kitaev chains.  }
    \label{fig:toyModel}
\end{figure}

A simple example with non-zero Berry phase can be constructed from a TRS analog of the 1D Kitaev chain \cite{Kitaev2001}, which is the simplest 1D model that supports Majorana fermions.
Kitaev's original model consists of spinless electrons living on a discrete 1D chain. In second quantization the electrons are described by complex fermionic operators
each of which may be decomposed into two hermitian Majorana operators.
In  Kitaev's original spinless model at the special point in parameter space where electron hopping
and $p$-wave pairing amplitude are equal and the chemical potential is zero, the model dimerizes in
a way that leaves a perfectly localized MBS at each end. To construct a time reversal invariant
analog, we take two copies of the Kitaev chain with opposite spin directions and opposite amplitude
for $p$-wave pairing, at the same special point in parameter space. This is sketched in the top
of Fig.~\ref{fig:toyModel}.
Our minimal model includes then, starting from one end of the wire,  exactly one MBS
Kramers pair and one Kramers pair of bulk states. This is illustrated in Fig.~\ref{fig:toyModel}.
The Hamiltonian for this system takes the form
\begin{equation}    \label{eqn:bulkHamiltonian}
    H_0= \frac {E_{\mbox{g}}} 2\sum_\sigma i\gamma_{2a\sigma}\gamma_{1b\sigma}
    = E_{\mbox{g}} \sum_{\sigma} d^\dagger_{\sigma}d_\sigma + \mbox{const.}
\end{equation}

The last expression is obtained by introducing the more familiar fermionic Bogoliubov
quasiparticle operators $d_{\sigma}=\tfrac 1 2(\gamma_{2a\sigma}+i\gamma_{1b\sigma})$.

For the two MBS to mix in a way that preserves TRS, we need  to couple them  to the bulk fermionic
mode. One coupling we use is a local change of the chemical potential: $ H_{\mu} =  \mu\sum_\sigma
c^\dagger_{1\sigma} c_{1\sigma}$. This coupling mixes the end point with the bulk, but does not mix
the two spin directions.  We expect that the different spin directions need to be coupled, to have
mixing of the Kramers pair. For this purpose we introduce $s$-wave pairing, $ H_{\Delta} =
\Delta(c_{1\uparrow} c_{1\downarrow} + c^\dagger_{1\downarrow} c^\dagger_{1\uparrow})$, on the last
site. We write the new terms in terms of Majorana operators to better understand which Majoranas
couple:
\begin{subequations}
\begin{align}
   H_{\mu} =& \frac \mu 2 \sum_\sigma\left(
   i\gamma_{1a\sigma}\gamma_{1b\sigma} + 1 \right),
   \label{eqn:chemicalPotentialCoupling}
   \\
   H_{\Delta} =& \frac \Delta 2 \left(
   i\gamma_{1a\uparrow}\gamma_{1b\downarrow} + i\gamma_{1b\uparrow}\gamma_{1a\downarrow}
   \right).
\end{align}
    \label{eqn:MajoranaBulkcoupling}
\end{subequations}
The full Hamiltonian of our toy model now reads is $H=H_0 + H_\mu + H_\Delta$ and
the couplings are illustrated in Fig.~\ref{fig:toyModel}.

The new Kramers pair of zero energy Majorana operators, in the presence of all couplings, is determined by the conditions $\left[ H,\gamma\right]=0$ and $\tilde\gamma=\TTT\gamma \TTT^{-1}$. To present the solutions more compactly, we change parameters:
\begin{subequations}
    \begin{align}
       \mu &= B\cos\alpha,
       \\
       \Delta &= B\sin\alpha.
    \end{align}
        \label{eqn:ParameterChange}
\end{subequations}
Here $\alpha$ parameterizes the ratio of bulk coupling between same spin and opposite spin.
Additionally we introduce $\tan\theta= B/ {E_{\mbox{g}}}$, which measures coupling between the
initial Majoranas and the bulk. With this notation, the results are
\begin{subequations}
\begin{align}
    \gamma(\theta,\alpha) & = \cos\theta \gamma_{1a\uparrow} - \sin\theta\left(
    \cos\alpha\gamma_{2a\uparrow} + \sin\alpha\gamma_{2a\downarrow} \right)\!,
     \\
    \tilde\gamma(\theta,\alpha)& = \cos\theta \gamma_{1a\downarrow} - \sin\theta\left(
    \cos\alpha\gamma_{2a\downarrow} - \sin\alpha\gamma_{2a\uparrow} \right)\!.
\end{align}
\label{eqn:MajoranaSolution}
\end{subequations}

Equation \eqref{eqn:MajoranaSolution} together with \eqref{eqn:BerryPhaseAndPotential} is used to calculate the Berry potential:
\begin{subequations}
\begin{align}
    A_\alpha &= -\frac 1 2 \sin^2\theta,
    \\
    A_\theta &= 0.
\end{align}
    \label{eqn:connectionExample}
\end{subequations}
For a loop in parameter space $\alpha:0\to 2\pi$, one thus gets the nontrivial contribution
$\varphi=\oint A_\alpha\mathrm{d}\alpha=-\pi\sin^2\theta$. Moreover, the Berry curvature is given by
$\Omega_{\theta\alpha}=-\sin\theta\cos\theta,$ which is non-zero. Note that for small $B$, the mixing
is proportional to $1/E_G^2$. Thus an external low frequency noise that leads to a fluctuating time
dependence of the Hamiltonian would lead to a decoherence time that grows only algebraically large
with $E_G$.

For the rest of this letter we will use BdG formalism, in which an operator is expressed as a four component spinor, and use the bra and ket notation to denote these spinors. The usual translation rules from
second quantized operators to BdG  states imply for Eq.~\eqref{eqn:BerryPhaseAndPotential} that
$\bm{A} =\tfrac 1 2\left\{ \gamma_{\bm{\eta}},\nabla_{\bm{\eta}} \tilde{\gamma}_{\bm{\eta}}
\right\}=\langle \gamma_{\bm{\eta}}|\nabla_{\bm{\eta}}|\tilde{\gamma}_{\bm\eta} \rangle$. The
factor of $\tfrac 1 2$ is a result of the Majorana states being normalized to 1, while the operators
anti-commute with themselves to 2.

As a second example for mixing we present numerical calculations of the Berry curvature
\cite{foot2} for a continuous 1D TRS $p$-wave superconductor. The Hamiltonian then reads,
\begin{equation}
        H=\left(\frac {p^2} {2m} -\mu(x)\right)\tau_z + p(\bm{\alpha}\cdot\bm{\sigma})\tau_z +
        p(\bm{v_}\Delta\cdot\bm{\sigma})\tau_x + \Delta\tau_x,
        \label{eqn:genralDIIIHamiltonianNoChannels}
\end{equation}
with an associated operator spinor
$\bm{\Psi}=(\Psi_{\uparrow},\Psi_{\downarrow},\Psi^\dagger_{\downarrow},-\Psi^\dagger_{\uparrow})^T$.
The symmetries of this Hamiltonian are
 $\PPP=\sigma_y\tau_yK$ and $\TTT=i\sigma_yK$, making DIII the relevant symmetry class.
The Hamiltonian describes a 1D system with $p$-wave pairing, $\bm{v}_\Delta$, $s$-wave pairing, $\Delta$, and spin-orbit
interaction, $\bm{\alpha}$. If $\bm{\alpha}=0$, $\Delta=0$ and $\bm{v_}\Delta=(v_x,0,0)^T$, this model
is the continuum version of two Kitaev wires with opposite spin and $p$-wave pairing.
If $\bm{\alpha}$ and $\Delta$ are non-zero, the two spin directions mix.

\begin{figure}
\centerline{\includegraphics[]{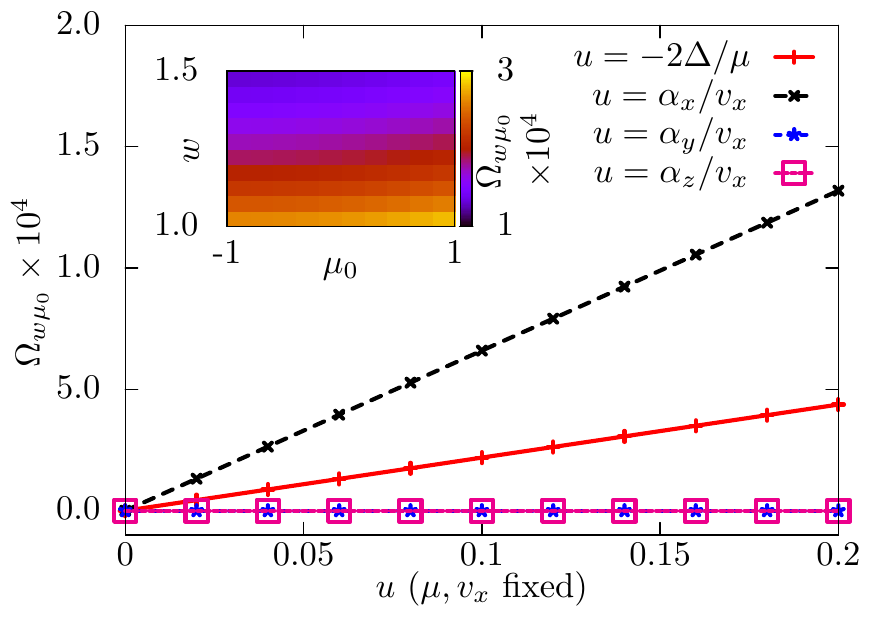}}
\caption{Berry curvature for the Hamiltonian \eqref{eqn:genralDIIIHamiltonianNoChannels}
as a function either $\Delta,\alpha_x,\alpha_y$, or $\alpha_z$, while the others of those four parameters are set to zero. The legend shows which parameter is varied in each case. The remaining parameters are $m=1$, $\mu=10$, $\mu_0=0$, $\bm{v}_\Delta=(5,0,0)^T$. The inset shows a larger area of the Berry curvature for the case $ {\alpha_x}/         {v_{\Delta}}=0.2$. The unit of length for all plots is 1. The total wire has a length of 40 and $x_0$ is         measured from the middle of the wire.}
    \label{fig:Berry}
\end{figure}

For certain parameter values the Hamiltonian \eqref{eqn:genralDIIIHamiltonianNoChannels} is in the
topological phase. This can then be controlled by the chemical potential, which means that the
Hamiltonian will be in the topological phase for $\mu>\mu_{\mbox{c}}$ and in the trivial
phase for $\mu<\mu_{\mbox{c}}$, where $\mu_{\mbox{c}}$ is determined by the other parameters in the
Hamiltonian.
We introduced a position dependent chemical potential of the form $\mu(x)=\mu_0+\mu\tanh(\tfrac {x-x_0} w)$,
that crosses from $\mu_0-\mu$ to $\mu_0+\mu$ around $x_0$. As long as $\mu_c$ is within this
interval, there will be a Kramers pair of MBS localized around $x_0$.
We will now study the adiabatic process involving this Kramers pair, where the parameters $w$ and $\mu_0$ change.
Note that both parameters enter the chemical potential which
does not couple spin directions. Figure \ref{fig:Berry} shows how a finite Berry
curvature is obtained if the Hamiltonian contains parameters that mix certain spin directions.
The main plot only shows the Berry curvature in one point,
which is of course not enough to calculate a Berry phase. The parameter dependence of the Berry
curvature is exemplified in the inset. This shows that there is a non-zero $\Omega$ in a whole
area.

As the last major part of the paper we show a sufficient condition for $\varphi=0$. Mixing cannot occur if the system can be
decomposed into two uncoupled one-dimensional subsystems that are time-reversed partners of one another and this decomposition can be done independently of
$\bm\eta$. If such a decomposition exists, it can be described by a non-singular hermitian operator $\Pi$. We require the operator to anti-commute with $\TTT$, such that
$\Pi$ will take opposite eigenvalues on the two subsystems. We further require it to commute with $\PPP$, which means that the MBSs can be chosen as eigenvectors of $\Pi$.

To formalize, if we find an operator $\Pi$ that satisfies
\begin{subequations}
\label{eqn:PIconditions}
\begin{align}
    [H(\bm{\eta}),\Pi]&= 0,
    \label{eqn:PiHCondition}\\
    [\PPP,\Pi]&=0,
    \label{eqn:PiPCondition}\\
    \{\TTT,\Pi\}&=0,
    \label{eqn:PiTCondition}
\end{align}
\end{subequations}
we can choose $\Pi | \gamma_{\bm{\eta}}\rangle=| \gamma_{\bm{\eta}}\rangle$ and $\Pi |\tilde
\gamma_{\bm\eta}\rangle=- | \tilde\gamma_{\bm\eta}\rangle$.  Then
$\nabla_{\bm\eta}|\tilde\gamma_{\bm{\eta}}\rangle$
is also an eigenstate of $\Pi$ with eigenvalue $-1$ and hence we can deduce that $\bm{A}=\langle
\gamma_{\bm\eta}|\nabla_{\bm{\eta}}|\tilde\gamma_{\bm\eta}\rangle=0$, and the two states are not mixed by the time
evolution of the Hamiltonian.

We now search for an operator $\Pi$ for the Hamiltonian \eqref{eqn:genralDIIIHamiltonianNoChannels} and the
corresponding $\TTT$ and $\PPP$ operators.
All operators of the form
$\Pi_{\bm{\hat{l}}}=\bm{\hat{l}}\cdot\bm{\sigma}\tau_z$ fulfill the conditions
\eqref{eqn:PiPCondition} and \eqref{eqn:PiTCondition} and commute with the kinetic energy part
    of $H$: $(\tfrac {p^2}{2m} - \mu)\tau_z$.
The commutators of $\Pi_{\bm{\hat l}}$ with the other terms of the
Hamiltonian also have to vanish, which constrains the vector
$\bm{\hat l}$. The constraints are
\begin{subequations}
\begin{align}
        0&=[H_{\bm{\alpha}},\Pi_{\bm{\hat{l}}}]=2i(\bm{\alpha}\times\bm{\hat{l}})\cdot \bm{\sigma},
        \\
        0&=\left[H_{\bm{v_\Delta}},\Pi_{\bm{\hat{l}}}\right]=
        -ip(\bm{v_\Delta}\cdot\bm{\hat{l}})\tau_y,
        \\
        0&=[H_\Delta,\Pi_{\bm{\hat{l}}}]=-2i\Delta(\bm{\hat{l}}\cdot \bm{\sigma})\tau_y.
\end{align}
\label{eqn:lConstraints}
\end{subequations}
The last constraint can never be met for any non-zero $\Delta$.
For the case $\Delta=0$, we get the conditions
$\bm{\alpha}\parallel{\bm{\hat l}} \perp \bm{v_\Delta}$, which can be fulfilled if $\bm{\alpha}
\perp\bm{v_\Delta}$. This agrees with our numerical results in Fig.~\ref{fig:Berry}, where the only
spin-orbit component generating a non-zero curvature is $\alpha_x$, which is parallel to $v_x$.
Thus, as long as $\Delta$ vanishes and $\bm{\alpha}$ is perpendicular to $\bm{v}_\Delta$, the Kramers pair
of MBS are not mixed.

The strategy we used above for finding $\Pi$ by first finding the basis for the space of hermitian operators that fulfill
\eqref{eqn:PiPCondition} and \eqref{eqn:PiTCondition} is useful also for more complicated systems. To show how this is done, we consider next a system where the spinor structure of our Hamiltonian is constructed out of tensor products of particle-hole Pauli matrices $\tau_i$, spin Pauli matrices $\sigma_i$ and matrices describing other orbital degrees of freedom denoted by $\lambda_i$. Furthermore, the $\lambda_i$-basis is chosen such that the individual elements are either fully real or imaginary, and we denote those matrices with $\lambda_i^R$ and  $\lambda_i^I$ accordingly. From \eqref{eqn:PiPCondition} and \eqref{eqn:PiTCondition} it follows that the basis matrices have to anti-commute with the chiral symmetry operator $\CCC=i\TTT\PPP=\tau_y$. Consequently we can choose them to be proportional to either $\tau_x$ or $\tau_z$. With the chiral symmetry condition fulfilled, we only need to fulfill one other of the conditions \eqref{eqn:PiPCondition} and \eqref{eqn:PiTCondition}, leaving the remaining one automatically fulfilled. We choose \eqref{eqn:PiTCondition} and in order to fulfill it our basis matrices have to be either proportional to $\sigma_i$ or fully imaginary, i.e. proportional to $\lambda^I_i$. In total we have the following allowed basis matrices from which we can construct candidates for $\Pi$
\begin{equation}
    \lambda^R_i\sigma_j\tau_z\mbox{, }\;\lambda^I_i\tau_z\mbox{, }\;\lambda^R_i\sigma_j\tau_x\mbox{,
    }\;\lambda^I_i\tau_x.
    \label{eqn:Pibasis}
\end{equation}

As an example we apply the procedure to the case of two separate spinful wires with intra and inter wire $s$-wave paring
\cite{Keselman2013,Gaidamauskas2014}. In this case, the orbital matrices $\lambda_i$ with $i=0\dots3$
are simply the three Pauli matrices (index 1 to 3) and the identity matrix (index 0). This means
that there is exactly one $\lambda^I=\lambda_y$ and all other orbital matrices are real.
We analyze a special case of the Hamiltonian from \cite{Gaidamauskas2014}:
\begin{equation}
    H=\left( \frac {p^2} {2m} -\mu + t\lambda_x +p\beta\lambda_z\sigma_z \right)\tau_z +
    \Delta_3\lambda_z\tau_x.
    \label{eqn:SpecialTwoWireHamiltonian}
\end{equation}
$\TTT$ and $\PPP$ are the same as the ones we used earlier.
In this case, the only possible symmetry operator is proportional to $\Pi=\lambda_x\sigma_z\tau_z$. The operator $\Pi$ is thus a combination of symmetries not related to frequently studied symmetries such as inversion or spin rotation. The reason for this is that the above conditions require $\Pi$ to commute with the particle-hole operator $\mathcal{P}$. Furthermore, note that $\Pi$ is proportional to $\lambda_x$ which means that the two one-dimensional subsystems, into which it decomposes the system, are not the two physical wires. Also note that the symmetry that allows the decomposition is easily broken, for example by $\lambda_x$ terms in the spin-orbit coupling or in the induced pairing.

In summary, we have shown how local mixing of a Kramers pair of Majorana fermions generates a transformation on the ground state of a DIII wire. This has implications for potential qubit applications. We have shown how adiabatic processes can generate such mixing and discussed an analytic toy model as well as numerical results as examples for such processes. Finally, we presented a symmetry condition that guarantees the absence of mixing. Designing systems with one or more of such symmetries (or approximate symmetries) can be helpful when attempting to minimize the mixing.

\textit{Acknowledgements} The Center for Quantum Devices is funded by the Danish National Research Foundation. The research was support by The Danish Council for Independent Research \textbar Natural Sciences. AS acknowledges support from the ERC, Minerva foundation, the US-Israel BSF and
Microsoft's Station Q. We thank E. Berg, A. Haim,  A. Keselman,  and Y. Oreg for discussions.


\begin{thebibliography}{16}%
\makeatletter
\providecommand \@ifxundefined [1]{%
 \@ifx{#1\undefined}
}%
\providecommand \@ifnum [1]{%
 \ifnum #1\expandafter \@firstoftwo
 \else \expandafter \@secondoftwo
 \fi
}%
\providecommand \@ifx [1]{%
 \ifx #1\expandafter \@firstoftwo
 \else \expandafter \@secondoftwo
 \fi
}%
\providecommand \natexlab [1]{#1}%
\providecommand \enquote  [1]{``#1''}%
\providecommand \bibnamefont  [1]{#1}%
\providecommand \bibfnamefont [1]{#1}%
\providecommand \citenamefont [1]{#1}%
\providecommand \href@noop [0]{\@secondoftwo}%
\providecommand \href [0]{\begingroup \@sanitize@url \@href}%
\providecommand \@href[1]{\@@startlink{#1}\@@href}%
\providecommand \@@href[1]{\endgroup#1\@@endlink}%
\providecommand \@sanitize@url [0]{\catcode `\\12\catcode `\$12\catcode
  `\&12\catcode `\#12\catcode `\^12\catcode `\_12\catcode `\%12\relax}%
\providecommand \@@startlink[1]{}%
\providecommand \@@endlink[0]{}%
\providecommand \url  [0]{\begingroup\@sanitize@url \@url }%
\providecommand \@url [1]{\endgroup\@href {#1}{\urlprefix }}%
\providecommand \urlprefix  [0]{URL }%
\providecommand \Eprint [0]{\href }%
\providecommand \doibase [0]{http://dx.doi.org/}%
\providecommand \selectlanguage [0]{\@gobble}%
\providecommand \bibinfo  [0]{\@secondoftwo}%
\providecommand \bibfield  [0]{\@secondoftwo}%
\providecommand \translation [1]{[#1]}%
\providecommand \BibitemOpen [0]{}%
\providecommand \bibitemStop [0]{}%
\providecommand \bibitemNoStop [0]{.\EOS\space}%
\providecommand \EOS [0]{\spacefactor3000\relax}%
\providecommand \BibitemShut  [1]{\csname bibitem#1\endcsname}%
\let\auto@bib@innerbib\@empty
\bibitem [{\citenamefont {Alicea}(2012)}]{AliceaReview}%
  \BibitemOpen
  \bibfield  {author} {\bibinfo {author} {\bibfnamefont {J.}~\bibnamefont
  {Alicea}},\ }\href {\doibase 10.1088/0034-4885/75/7/076501} {\bibfield
  {journal} {\bibinfo  {journal} {Rep.~Prog.~Phys.}\ }\textbf {\bibinfo
  {volume} {75}},\ \bibinfo {pages} {076501} (\bibinfo {year}
  {2012})}\BibitemShut {NoStop}%
\bibitem [{Bee()}]{BeenakkerReview}%
  \BibitemOpen
  \href@noop {} {}\bibinfo {note} {C. W. J. Beenakker, Annu. Rev. Condens.
  Matter Phys. 4, 113 (2013)}\BibitemShut {NoStop}%
\bibitem [{\citenamefont {Leijnse}\ and\ \citenamefont
  {Flensberg}(2012)}]{LeijnseReview}%
  \BibitemOpen
  \bibfield  {author} {\bibinfo {author} {\bibfnamefont {M.}~\bibnamefont
  {Leijnse}}\ and\ \bibinfo {author} {\bibfnamefont {K.}~\bibnamefont
  {Flensberg}},\ }\href {\doibase 10.1088/0268-1242/27/12/124003} {\bibfield
  {journal} {\bibinfo  {journal} {Semiconductor Science and Technology}\
  }\textbf {\bibinfo {volume} {27}},\ \bibinfo {pages} {124003} (\bibinfo
  {year} {2012})}\BibitemShut {NoStop}%
\bibitem [{\citenamefont {Nayak}\ \emph {et~al.}(2008)\citenamefont {Nayak},
  \citenamefont {Simon}, \citenamefont {Stern}, \citenamefont {Freedman},\ and\
  \citenamefont {{Das Sarma}}}]{Nayak2008}%
  \BibitemOpen
  \bibfield  {author} {\bibinfo {author} {\bibfnamefont {C.}~\bibnamefont
  {Nayak}}, \bibinfo {author} {\bibfnamefont {S.}~\bibnamefont {Simon}},
  \bibinfo {author} {\bibfnamefont {A.}~\bibnamefont {Stern}}, \bibinfo
  {author} {\bibfnamefont {M.}~\bibnamefont {Freedman}}, \ and\ \bibinfo
  {author} {\bibfnamefont {S.}~\bibnamefont {{Das Sarma}}},\ }\href {\doibase
  10.1103/RevModPhys.80.1083} {\bibfield  {journal} {\bibinfo  {journal}
  {Rev.~Mod.~Phys.}\ }\textbf {\bibinfo {volume} {80}},\ \bibinfo {pages}
  {1083} (\bibinfo {year} {2008})}\BibitemShut {NoStop}%
\bibitem [{\citenamefont {Wong}\ and\ \citenamefont {Law}(2012)}]{Wong2012}%
  \BibitemOpen
  \bibfield  {author} {\bibinfo {author} {\bibfnamefont {C.~L.~M.}\
  \bibnamefont {Wong}}\ and\ \bibinfo {author} {\bibfnamefont {K.~T.}\
  \bibnamefont {Law}},\ }\href {\doibase 10.1103/PhysRevB.86.184516} {\bibfield
   {journal} {\bibinfo  {journal} {Phys.~Rev.~B}\ }\textbf {\bibinfo {volume}
  {86}},\ \bibinfo {pages} {184516} (\bibinfo {year} {2012})}\BibitemShut
  {NoStop}%
\bibitem [{\citenamefont {Zhang}\ \emph {et~al.}(2013)\citenamefont {Zhang},
  \citenamefont {Kane},\ and\ \citenamefont {Mele}}]{Zhang2013}%
  \BibitemOpen
  \bibfield  {author} {\bibinfo {author} {\bibfnamefont {F.}~\bibnamefont
  {Zhang}}, \bibinfo {author} {\bibfnamefont {C.~L.}\ \bibnamefont {Kane}}, \
  and\ \bibinfo {author} {\bibfnamefont {E.~J.}\ \bibnamefont {Mele}},\ }\href
  {\doibase 10.1103/PhysRevLett.111.056402} {\bibfield  {journal} {\bibinfo
  {journal} {Phys.~Rev.~Lett.}\ }\textbf {\bibinfo {volume} {111}},\ \bibinfo
  {pages} {056402} (\bibinfo {year} {2013})}\BibitemShut {NoStop}%
\bibitem [{\citenamefont {Nakosai}\ \emph {et~al.}(2013)\citenamefont
  {Nakosai}, \citenamefont {Budich}, \citenamefont {Tanaka}, \citenamefont
  {Trauzettel},\ and\ \citenamefont {Nagaosa}}]{Nakosai2013}%
  \BibitemOpen
  \bibfield  {author} {\bibinfo {author} {\bibfnamefont {S.}~\bibnamefont
  {Nakosai}}, \bibinfo {author} {\bibfnamefont {J.~C.}\ \bibnamefont {Budich}},
  \bibinfo {author} {\bibfnamefont {Y.}~\bibnamefont {Tanaka}}, \bibinfo
  {author} {\bibfnamefont {B.}~\bibnamefont {Trauzettel}}, \ and\ \bibinfo
  {author} {\bibfnamefont {N.}~\bibnamefont {Nagaosa}},\ }\href@noop {}
  {\bibfield  {journal} {\bibinfo  {journal} {Phys.~Rev.~Lett.}\ }\textbf
  {\bibinfo {volume} {110}},\ \bibinfo {pages} {117002} (\bibinfo {year}
  {2013})}\BibitemShut {NoStop}%
\bibitem [{\citenamefont {Nakosai}\ \emph {et~al.}(2012)\citenamefont
  {Nakosai}, \citenamefont {Tanaka},\ and\ \citenamefont
  {Nagaosa}}]{Nakosai2012}%
  \BibitemOpen
  \bibfield  {author} {\bibinfo {author} {\bibfnamefont {S.}~\bibnamefont
  {Nakosai}}, \bibinfo {author} {\bibfnamefont {Y.}~\bibnamefont {Tanaka}}, \
  and\ \bibinfo {author} {\bibfnamefont {N.}~\bibnamefont {Nagaosa}},\ }\href
  {\doibase 10.1103/PhysRevLett.108.147003} {\bibfield  {journal} {\bibinfo
  {journal} {Phys.~Rev.~Lett.}\ }\textbf {\bibinfo {volume} {108}},\ \bibinfo
  {pages} {147003} (\bibinfo {year} {2012})}\BibitemShut {NoStop}%
\bibitem [{\citenamefont {Deng}\ \emph {et~al.}(2012)\citenamefont {Deng},
  \citenamefont {Viola},\ and\ \citenamefont {Ortiz}}]{Deng2012}%
  \BibitemOpen
  \bibfield  {author} {\bibinfo {author} {\bibfnamefont {S.}~\bibnamefont
  {Deng}}, \bibinfo {author} {\bibfnamefont {L.}~\bibnamefont {Viola}}, \ and\
  \bibinfo {author} {\bibfnamefont {G.}~\bibnamefont {Ortiz}},\ }\href@noop {}
  {\bibfield  {journal} {\bibinfo  {journal} {Phys. \ Rev. \ Lett.}\ }\textbf
  {\bibinfo {volume} {108}},\ \bibinfo {pages} {036803} (\bibinfo {year}
  {2012})}\BibitemShut {NoStop}%
\bibitem [{\citenamefont {Keselman}\ \emph {et~al.}(2013)\citenamefont
  {Keselman}, \citenamefont {Fu}, \citenamefont {Stern},\ and\ \citenamefont
  {Berg}}]{Keselman2013}%
  \BibitemOpen
  \bibfield  {author} {\bibinfo {author} {\bibfnamefont {A.}~\bibnamefont
  {Keselman}}, \bibinfo {author} {\bibfnamefont {L.}~\bibnamefont {Fu}},
  \bibinfo {author} {\bibfnamefont {A.}~\bibnamefont {Stern}}, \ and\ \bibinfo
  {author} {\bibfnamefont {E.}~\bibnamefont {Berg}},\ }\href {\doibase
  10.1103/PhysRevLett.111.116402} {\bibfield  {journal} {\bibinfo  {journal}
  {Phys.~Rev.~Lett.}\ }\textbf {\bibinfo {volume} {111}},\ \bibinfo {pages}
  {116402} (\bibinfo {year} {2013})}\BibitemShut {NoStop}%
\bibitem [{\citenamefont {Gaidamauskas}\ \emph {et~al.}(2014)\citenamefont
  {Gaidamauskas}, \citenamefont {Paaske},\ and\ \citenamefont
  {Flensberg}}]{Gaidamauskas2014}%
  \BibitemOpen
  \bibfield  {author} {\bibinfo {author} {\bibfnamefont {E.}~\bibnamefont
  {Gaidamauskas}}, \bibinfo {author} {\bibfnamefont {J.}~\bibnamefont
  {Paaske}}, \ and\ \bibinfo {author} {\bibfnamefont {K.}~\bibnamefont
  {Flensberg}},\ }\href {\doibase 10.1103/PhysRevLett.112.126402} {\bibfield
  {journal} {\bibinfo  {journal} {Phys.~Rev.~Lett.}\ }\textbf {\bibinfo
  {volume} {112}},\ \bibinfo {pages} {126402} (\bibinfo {year}
  {2014})}\BibitemShut {NoStop}%
\bibitem [{Hai()}]{Haim2013}%
  \BibitemOpen
  \href@noop {} {}\bibinfo {note} {A. Haim, A. Keselman, E. Berg, Y. Oreg,
  arXiv:1310.4525}\BibitemShut {NoStop}%
\bibitem [{\citenamefont {Liu}\ \emph {et~al.}(2014)\citenamefont {Liu},
  \citenamefont {Wong},\ and\ \citenamefont {Law}}]{Liu2014}%
  \BibitemOpen
  \bibfield  {author} {\bibinfo {author} {\bibfnamefont {X.-J.}\ \bibnamefont
  {Liu}}, \bibinfo {author} {\bibfnamefont {C.~L.~M.}\ \bibnamefont {Wong}}, \
  and\ \bibinfo {author} {\bibfnamefont {K.~T.}\ \bibnamefont {Law}},\ }\href
  {\doibase 10.1103/PhysRevX.4.021018} {\bibfield  {journal} {\bibinfo
  {journal} {Phys.~Rev.~X}\ }\textbf {\bibinfo {volume} {4}},\ \bibinfo {pages}
  {021018} (\bibinfo {year} {2014})}\BibitemShut {NoStop}%
\bibitem [{foo({\natexlab{a}})}]{foot1}%
  \BibitemOpen
  \href@noop {} {} ({\natexlab{a}}),\ \bibinfo {note} {identities, which are
  obtained by differentiating $\left\{
  \gamma_{\bm{\eta}(t)},\gamma_{\bm{\eta}(t)} \right\}=2$, $\left\{
  \tilde\gamma_{\bm{\eta}(t)},\tilde\gamma_{\bm{\eta}(t)} \right\}=2$ and
  $\left\{ \gamma_{\bm{\eta}(t)},\tilde\gamma_{\bm{\eta}(t)} \right\}=0$, are
  used to obtain the result.}\BibitemShut {Stop}%
\bibitem [{\citenamefont {Kitaev}(2001)}]{Kitaev2001}%
  \BibitemOpen
  \bibfield  {author} {\bibinfo {author} {\bibfnamefont {A.~Y.}\ \bibnamefont
  {Kitaev}},\ }\href {\doibase 10.1070/1063-7869/44/10S/S29} {\bibfield
  {journal} {\bibinfo  {journal} {Phys.~Usp.}\ }\textbf {\bibinfo {volume}
  {44}},\ \bibinfo {pages} {131} (\bibinfo {year} {2001})}\BibitemShut
  {NoStop}%
\bibitem [{foo({\natexlab{b}})}]{foot2}%
  \BibitemOpen
  \href@noop {} {} ({\natexlab{b}}),\ \bibinfo {note} {we use the formula
  $\Omega_{\eta_i\eta_j}= \sum_{n}\tfrac {1} {E_n^2}
  [\protect\langle\gamma|\partial_{\eta_i} H|n \protect\rangle\protect\langle
  n|\partial_{\eta_j} H|\tilde\gamma \protect\rangle -
  \protect\langle\gamma|\partial_{\eta_j} H|n\protect\rangle\protect\langle
  n|\partial_{\eta_i} H|\tilde\gamma \protect\rangle]$, where the sum is over
  all the bulk states.}\BibitemShut {Stop}%
\end{thebibliography}

%

\end{document}